# Energy-Arena: A Dynamic Benchmark for Operational Energy Forecasting

Max Kleinebrahm[1], Jonathan Berrisch[2], Philipp Eiser[1], Wolf Fichtner[1], Veit Hagenmeyer[3], Matthias Hertel[3], Nils Koster[4], Sebastian Lerch[5,6], Ralf Mikut[3], Jan Priesmann[7], Melanie Schienle[4,5], Benjamin Schäfer[3], Jann Weinand[8], Florian Ziel[2]

[1] Karlsruhe Institute of Technology, Institute for Industrial Production, Karlsruhe, Germany

[2] University of Duisburg-Essen, House of Energy Markets and Finance, Essen, Germany

[3] Karlsruhe Institute of Technology, Institute for Automation and Applied Informatics, Eggenstein-Leopoldshafen, Germany

[4] Karlsruhe Institute of Technology, Institute of Statistics, Karlsruhe, Germany

[5] Heidelberg Institute for Theoretical Studies, Heidelberg, Germany

[6] Marburg University, Department of Mathematics and Computer Science, Marburg, Germany

[7] Volatile GmbH, Cologne, Germany

[8] Forschungszentrum Jülich GmbH, Institute of Climate and Energy Systems – Jülich Systems Analysis, Jülich, Germany

***Abstract* — Energy forecasting research faces a persistent comparability gap that makes it difficult to measure consistent progress over time. Reported accuracy gains are often not directly comparable because models are evaluated under study-specific datasets, time periods, information sets, and scoring setups, while widely used benchmarks and competition datasets are typically tied to fixed historical windows. This paper introduces the Energy-Arena, a dynamic benchmarking platform for operational energy time series forecasting that provides a continuously updated reference point as energy systems evolve. The platform operates as an open, API-based submission system and standardizes challenge definitions and submission deadlines aligned with operational constraints. Performance is reported on rolling evaluation windows via persistent leaderboards. By moving from retrospective backtesting to forward-looking benchmarking, the Energy-Arena enforces standardized ex-ante submission and ex-post evaluation, thereby improving transparency by preventing information leakage and retroactive tuning. The platform is publicly available at Energy-Arena.org.**

***Index Terms* — Benchmarking, Electricity price forecasting, Energy forecasting, Load forecasting, Renewable energy forecasting**

## I. Introduction

Forecasting electricity prices, load, and renewable generation has become essential for power system operators, energy traders, and analysts as renewable penetration and market volatility rise across Europe, a trend reflected in the sharp growth of related research, including more than 2,400 papers published in 2025 alone (see Fig. 1).

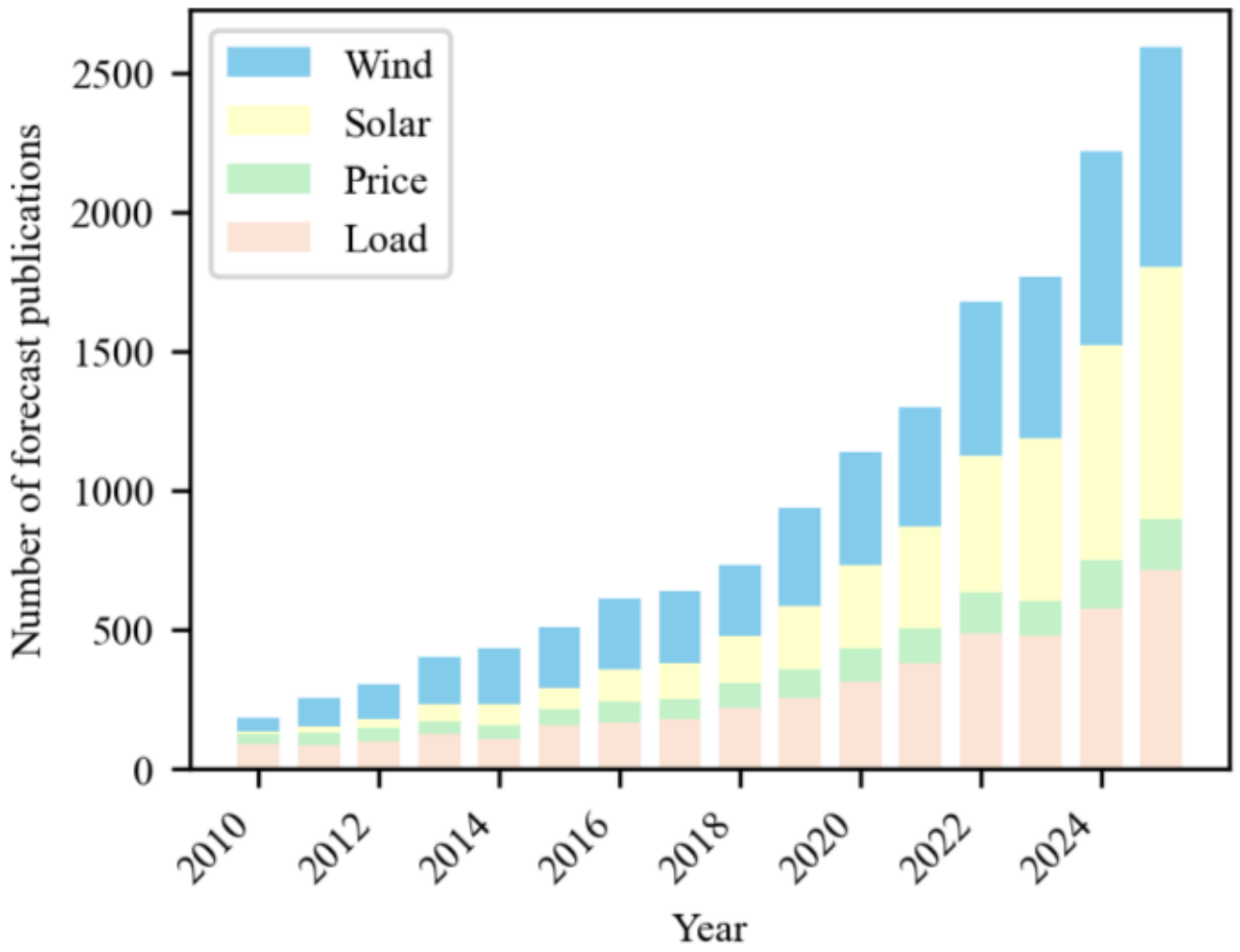

Figure 1. Increase in the annual number of energy forecasting publications indexed in Scopus from 2010 to 2025 (similar to [3]). The analysis is based on TITLE-ABS-KEY searches using a symmetric proximity structure: Wind: (("wind" W/3 ("power" OR "generation")) W/3 (forecast* OR predict*)); Solar: ((("solar" OR "photovoltaic" OR "PV") W/3 ("power" OR "generation")) W/3 (forecast* OR predict*)); Price: ((("electricity" OR "power") W/3 "price") W/3 (forecast* OR predict*)); Load: ((("electricity" OR "power") W/3 ("load" OR "demand")) W/3 (forecast* OR predict*)). W/3 denotes a proximity operator requiring that the specified terms occur within three words of each other in the given order. All queries were restricted to document types Article or Review (DOCTYPE(ar OR re)). Code for figure reproduction.

Despite a vast body of literature and notable open-access efforts e.g., Refs. [1], [2], the field lacks a widely accepted, continuously updated dynamic benchmark that captures the evolving conditions of modern energy systems. As a result, it remains difficult to assess state-of-the-art forecasting performance, hindering consistent and measurable progress, both in research and in commercial practice.

Most research papers rely on study-specific benchmark datasets, evaluation metrics, and experimental designs, which enable intra-study model comparisons but limit inter-study comparability. Even when addressing the same target variable, such as day-ahead electricity prices, models are evaluated on different time periods, market zones, preprocessing pipelines, and information sets [4]. As a result, reported performance differences often reflect experimental design choices rather than methodological advances. In addition, model comparisons may be affected by implementation choices and tuning effort. In practice, competing methods are not always implemented or optimized with equal care, for example with respect to feature engineering or hyperparameter selection. As noted by Hewamalage *et al.* [5], this can lead to misleading conclusions when state-of-the-art methods are compared under suboptimal configurations. As a result, reported performance differences may reflect differences in implementation quality rather than inherent methodological advantages.

Important efforts have been made to address this issue. In particular, Lago *et al.* [1] introduce a comprehensive open-access electricity price forecasting benchmark dataset covering five electricity markets, six years of data per market, standardized training and testing splits, and clearly defined evaluation protocols. Their framework substantially improves transparency and comparability by ensuring multi-market coverage, sufficiently long out-of-sample periods, and the inclusion of relevant day-ahead exogenous forecasts. Nevertheless, such benchmarks remain inherently static: once published, the dataset reflects a fixed historical period. In rapidly evolving energy systems characterized by increasing renewable penetration, structural shifts in bidding behavior, and regulatory changes, forecasting performance is highly time-dependent. A method that performs best on data from 2011–2018 may not remain competitive under current market conditions.

Beyond dataset heterogeneity, experimental designs are often underspecified with respect to the information structure assumed at forecast issuance. Some studies implicitly assume access to exogenous regressors that would not have been publicly available at gate closure. For example [6], [7], [8] use wind and solar generation forecasts from the European Network of Transmission System Operators for Electricity (ENTSO-E) [9] as input features for day-ahead price prediction. However, the publicly available day-ahead wind and solar forecasts are released only after 18:00 on the day before delivery, whereas the day-ahead market already closes at 12:00 on that day [9]. Their use therefore presumes that comparable wind and solar forecasts are available to market participants before gate closure, although this assumption is typically not formalized explicitly. This highlights the broader distinction between operational (ex-ante) forecasting models, which must rely only on information available at the forecast issuance time, and ex-post explanatory models that may incorporate variables observed later. As the publicly available ENTSO-E forecasts are released only after the auction, studies relying on these values cannot be reproduced under a strictly public operational information set. While market participants may indeed have access to proprietary or in-house forecasts prior to 12:00, the absence of an explicitly defined information cutoff limits operational reproducibility and comparability across studies. This limitation is further compounded by the fact that forecasting performance may depend on access to proprietary data sources, which are rarely disclosed explicitly, making it difficult to assess and compare the effective information sets underlying competing models.

More generally, the forecasting objective itself is sometimes insufficiently defined. Some studies appear to model prices predominantly as a function of fundamental drivers, implicitly excluding cross-market information such as earlier clearing results (e.g., EXAA prices) without explicitly mentioning this design choice. Yet such signals are available prior to EPEX day-ahead gate closure and are known to contain valuable information [7], [10]. The absence of a clearly stated information cutoff and modeling objective, whether fundamental explanation, statistical accuracy under operational constraints, or structural interpretation, further complicates cross-study comparison.

Forecasting competitions, such as the Global Energy Forecasting Competition [2], partially address this issue by imposing unified experimental conditions: a common dataset, predefined training and testing splits, standardized forecast horizons, and fixed evaluation metrics. These settings enhance transparency and ensure that model comparisons are conducted under identical framework conditions. However, competition datasets are inherently static and limited to a predefined historical window. As a result, while they resolve inconsistencies in experimental design within a given challenge period, they do not provide a continuously updated benchmark capable of assessing model robustness under evolving market regimes, changing renewable penetration levels, and shifting regulatory environments.

Taken together, heterogeneous datasets, inconsistent information assumptions, partially implicit modeling objectives, and the static nature of existing competitions create a persistent comparability gap in energy forecasting research. While commercial forecasting tools operate under operational constraints, they remain largely closed-source and non-transparent, preventing independent verification of performance claims. Consequently, neither academic backtests nor proprietary solutions currently provide a continuously updated, operationally consistent, and publicly comparable benchmark (see Tab. 1).

Dynamic forecasting competition platforms can reduce the comparability gap by moving evaluation from retrospective backtests to forward-looking assessment with standardized rules, automated scoring, and persistent leaderboards. Early examples include FlexUp Teamcast [11], which benchmarks short-term power market price forecasts, and TS-Arena [12], which provides leakage-free live evaluation of time-series foundation models on future data. In the broader machine learning and forecasting community, several recent benchmarks evaluate forecasting models across diverse domains, including ForecastBench [13] and fev-bench [14].

Particularly for foundation models, limited transparency regarding training data and potential overlap with benchmark datasets can complicate the interpretation of reported performance. While these initiatives improve standardization and task diversity, they are not specifically designed for operational energy forecasting: fev-bench relies on static historical datasets, and ForecastBench focuses on event-based prediction questions rather than structured time-series forecasting under operational constraints.

TABLE I. Stylized qualitative comparison of common energy forecast evaluation settings.

| Energy Forecasts | Transparent | Continuous | Comparable | Operational |
|---|---|---|---|---|
| Research paper | + | - | - | - |
| Commercial tools | - | + | - | + |
| Static competition | + | - | + | + |
| Dynamic competition | + | + | + | + |

This paper introduces the Energy-Arena, a dynamic benchmarking platform for energy time series forecasting. The platform establishes a shared reference point for evaluating forecasting approaches as energy systems evolve over time. By enabling continuous comparison of models under common evaluation conditions, the Energy-Arena reduces the comparability gap across studies and forecasting tools. For researchers, this facilitates more cumulative scientific progress by making methodological improvements easier to identify and validate. For practitioners and commercial providers, the platform provides a transparent environment for benchmarking forecasting performance against state-of-the-art approaches under real-world market dynamics. In this way, the Energy-Arena aims to create a continuously evolving benchmark that supports transparent and systematic progress in energy forecasting research.

## II. Energy-Arena

The following section first presents the overall platform architecture, followed by a description of the participation workflow. Finally, it outlines how challenges are defined within the Energy-Arena and how continuous forecasting progress is tracked through leaderboards and forecast visualizations.

### A. *Energy-Arena Platform Architecture*

Fig. 2 provides a schematic overview of the platform architecture. The Energy-Arena is a modular, Application Programming Interface (API)-driven benchmarking system that integrates standardized challenge definitions, participant interaction, automated scoring, and continuously updated leaderboards. Forecasting challenges are defined through configuration files specifying the challenge-specific framework conditions, such as the target variable and the temporal structure.

Participants interact with the platform through a web-based frontend that provides challenge descriptions, leaderboards, forecast visualizations, and a personalized dashboard with API key management. Forecast submissions are sent programmatically to the backend through a dedicated submission API. The backend validates incoming forecasts against the challenge configuration, enforces submission deadlines, e.g., corresponding to market gate-closure times, and stores participants, submissions, ground-truth values, and scoring results in a relational database. In addition to participant submissions, the platform can host a set of reference benchmark models to provide performance baselines. These may include simple statistical benchmarks (e.g., persistence or seasonal naïve forecasts) as well as more advanced machine learning models.

Ground-truth data are retrieved from external data providers (e.g., ENTSO-E [15]) and processed through a worker pipeline that periodically ingests new observations, evaluates pending submissions once realized values become available, and updates leaderboard aggregates over predefined evaluation windows. Leaderboards and forecast visualizations are then displayed in the frontend interface. Email services support account registration and verification workflows. Through this architecture, the Energy-Arena enables automated forward evaluation of forecasting models under explicitly defined submission deadlines and standardized scoring rules, providing a transparent and continuously updated benchmark based on newly realized energy system data.

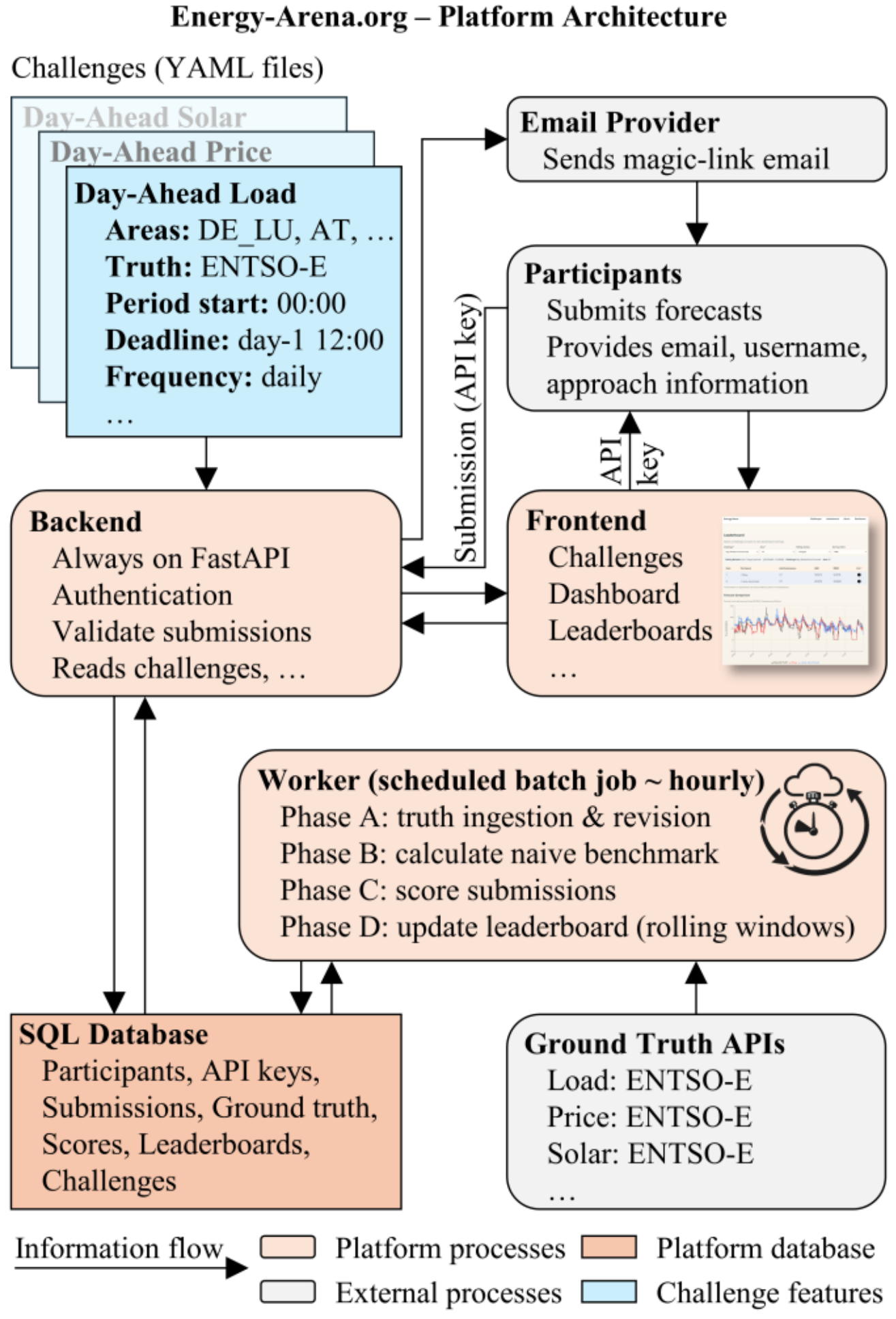

Figure 2. Schematic overview of the Energy-Arena platform architecture, illustrating the main system components and their interactions. (API = Application Programming Interface, ENTSO-E = European Network of Transmission System Operators for Electricity)

## B. *Participation Workflow*

Fig. 3 illustrates the participation workflow of the Energy-Arena. Participation consists of a one-time account setup followed by recurring automated forecast submissions aligned with the schedule of each forecasting challenge. Through the user dashboard, participants can generate API keys for programmatic submission and optionally provide metadata on their forecasting approach, such as a short method description, a code repository, or a link to a commercial service. They may also choose whether their submitted forecast trajectories are publicly visualized or whether only aggregated performance metrics are displayed.

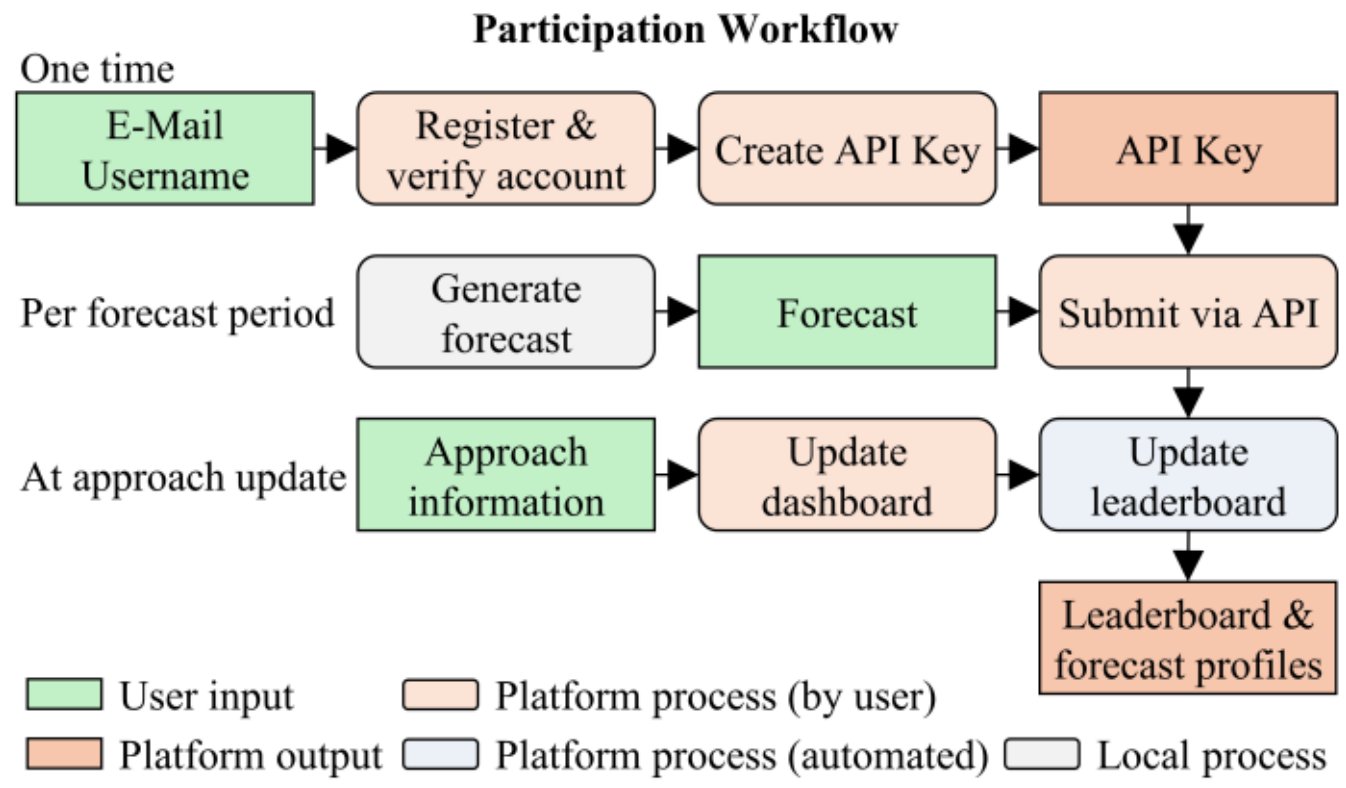

Figure 3. Visualization of the user participation workflow. (API = Application Programming Interface).

## C. *Challenge Design and Leaderboards*

Forecasting challenges in the Energy-Arena are defined through structured configuration files written in YAML, which specify all parameters required for data ingestion, submission validation, scoring, and leaderboard generation. This configuration-based approach allows new challenges to be added easily by extending the challenge repository with additional files, without requiring modifications to the platform's core logic. Each challenge definition includes the target variable, geographic areas, reference timezone, forecast horizon, submission deadlines and frequency, evaluation metrics, and leaderboard aggregation settings. Furthermore, the configuration specifies the data source for ground-truth observations (e.g., ENTSO-E Transparency Platform [15]), as well as constraints on forecast payloads such as timestamp structure, resolution, and value ranges.

The temporal structure of a forecasting challenge is illustrated in Fig. 4 using the example of a day-ahead load forecasting task. Although day-ahead forecasting is a common operational setup in electricity markets, the Energy-Arena framework is not restricted to this horizon and can support challenges with different forecast horizons and submission frequencies. In the example shown, forecasts must be submitted before a predefined gate-closure time (e.g., 12:00 CET on the day preceding delivery) and must cover the complete target period of the following calendar day in the specified reference timezone. Challenges are executed on a recurring schedule (e.g., daily cadence), generating a sequence of forecast targets over time.

In addition to target definitions and submission constraints, each challenge implicitly defines an operational information set through the submission deadline. Participants are free to use any data that would be available at the time of forecast issuance, including proprietary inputs such as weather forecasts or internally generated features. The platform does not enforce restrictions on exogenous variables, as such constraints would be difficult to verify in practice. Instead, participants may optionally provide metadata describing their input data sources (e.g., public vs. proprietary), enabling leaderboard segmentation and facilitating transparent comparison across different information regimes.

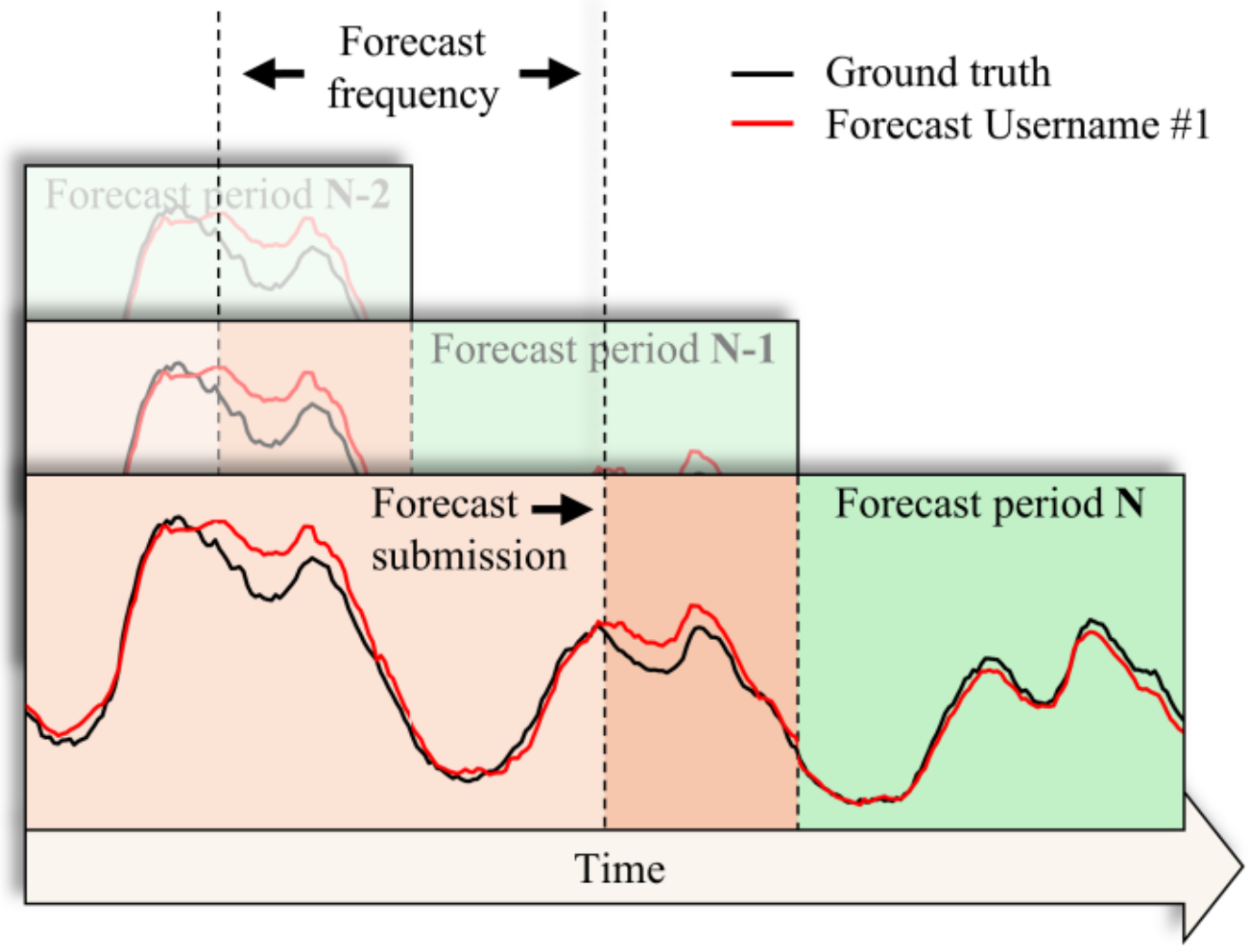

Figure 4. Exemplary visualization of the temporal relationship between forecast submission deadline, forecast period, and the frequency of forecasting events within the day-ahead load forecasting challenge.

Once the corresponding ground-truth values become available, submitted forecasts are automatically evaluated using predefined performance metrics such as Mean Absolute Error (MAE) and Root Mean Square Error (RMSE). These metrics are part of a standardized set provided by the platform and can be selected within the challenge configuration. In addition to point forecasts, the platform supports probabilistic submissions, including quantile and ensemble forecasts, which are evaluated using proper scoring rules such as the Continuous Ranked Probability Score (CRPS) and the Weighted Interval Score (WIS). Submissions may include point predictions as well as distributional information (e.g., sets of quantiles and/or ensemble members), allowing multiple forecast representations to be evaluated within a unified framework. The resulting scores are aggregated over rolling evaluation windows (e.g., 1, 7, 30, 90, and 365 days) and displayed on challenge leaderboards, enabling continuous comparison of forecasting approaches under consistent operational conditions. Fig. 5 illustrates an exemplary leaderboard for the day-ahead load forecasting challenge described above. Platform users can filter leaderboards by challenge, geographic area, and evaluation window, and sort results according to their preferred performance metric.

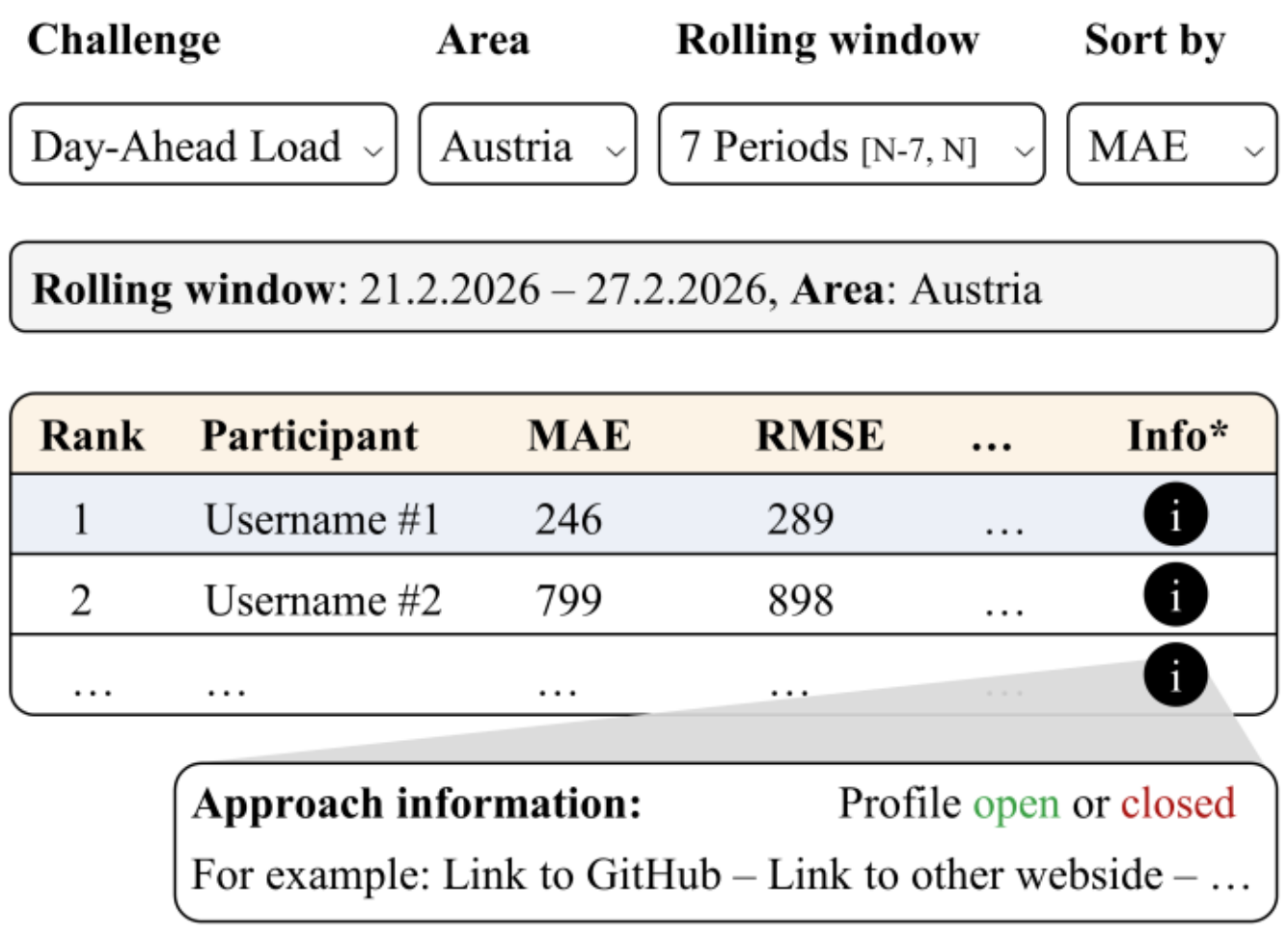

Figure 5. Exemplary visualization of the leaderboard for the day-ahead load forecasting challenge. Metrics are computed over the last N periods (here, N = 7) for which ground truth is available. (MAE = Mean Absolute Error, RMSE = Root Mean Square Error).

By selecting the information button in the final column, users can access additional details about the forecasting approach used by the respective participant, provided that such information has been supplied for the corresponding challenge–area combination via the dashboard. Furthermore, by selecting a participant entry, users can inspect the submitted forecast time series in an interactive visualization and compare forecasts across participants as well as against the realized ground truth. The visualization of forecast trajectories is only available if participants choose to make their forecasts publicly visible through the dashboard settings.

## III. Conclusion And Outlook

This paper introduced the Energy-Arena, a dynamic benchmarking platform designed to enable continuous and operationally consistent evaluation of energy forecasting models. The platform addresses a key limitation of existing research benchmarks and forecasting competitions, namely their static nature and limited reproducibility with respect to real-world information availability. By combining standardized challenge definitions, automated forward evaluation, and continuously updated leaderboards, the Energy-Arena provides a transparent reference point for comparing forecasting approaches under evolving energy system conditions.

Future development of the platform will focus on expanding the set of forecasting challenges and evaluation settings. While the current implementation primarily targets deterministic day-ahead forecasting tasks, future challenges may include probabilistic forecasts and scenario-based predictions for various energy time series. In addition, new geographic regions and temporal resolutions can be integrated through the configuration-based challenge design, allowing the platform to adapt to emerging research and industry needs. As an optional community-building mechanism, the platform could host time-limited themed competitions (e.g., seasonal or event-focused challenges) with standardized rules and optional prizes, while keeping the underlying challenges and leaderboards as a persistent, continuously updated benchmark.

Beyond methodological benchmarking, the Energy-Arena aims to foster collaboration between academia and industry. Researchers can use the platform to evaluate new forecasting methods under realistic operational constraints, while commercial providers can benchmark proprietary models against publicly available approaches. If participants choose to make their forecasts publicly visible, the submitted predictions can form a continuously growing archive of historical forecasts. Such an archive would enable the community to reconstruct the forecasting landscape at specific points in time and perform backtesting studies based on forecasts that were actually issued before the realization of the corresponding outcomes. The platform can also support educational applications, for example, through university forecasting challenges, such as [16], in which students develop and deploy forecasting models on the platform. In this way, the Energy-Arena aims to contribute to a more transparent, collaborative, and continuously evolving ecosystem for energy forecasting research and practice.

## IV. Author Contributions

M. Kleinebrahm conceived and designed the Energy-Arena platform, developed the benchmarking framework, implemented the system, and wrote the manuscript. F. Ziel, J. Berrisch, and N. Koster contributed to the conceptualization of the platform. All other authors contributed to reviewing and editing the manuscript.